\documentclass[epj,twocolumn]{webofc}
\usepackage[varg]{txfonts}   
\usepackage{xspace}
\newcommand{\gev}{\,\mathrm{GeV}} 
\newcommand{\mev}{\,\mathrm{MeV}} 
\newcommand{\compass}{\textsc{Compass}\xspace}

\usepackage[varg]{txfonts}   
\woctitle{Dark Matter, Hadron Physics and Fusion Physics}
\begin{document}
\title{Highlights from \textsc{Compass} in hadron spectroscopy}
%
%

\author{Fabian Krinner for the \textsc{Compass} collaboration\inst{1}\fnsep\thanks{\email{fabian-krinner@mytum.de}}
}

\institute{Technische Universit\"at M\"unchen, Physik-Department, E18
          }

\abstract{%
Since Quantum Choromdynamics allows for gluon self-coupling,
quarks and gluons cannot be observed as free particles, but only
their bound states, the hadrons. This so-called confinement phenomenon is
responsible for $98\%$ of the mass in the visible universe. The measurement of
the hadron excitation spectra therefore gives valuable input for
theory and phenomenology to quantitatively understand this phenomenon.	


One simple model to describe hadrons is the Constituent Quark Model (CQM), which
knows two types of hadrons: mesons, consisting of a quark and an antiquark, and baryons, 
which are made out of three quarks. More advanced models, which are inspired by QCD
as well as calculations within Lattice QCD predict the existence of other types of 
hadrons, which may be e.g. described solely by gluonic excitations (glueballs) or
mixed quark and gluon excitations (hybrids).


In order to search for such states, the \textsc{Compass} experiment at the Super Proton Synchrotron at CERN 
has collected large data sets, which allow to study the light-quark meson and baryon spectra in unmatched precision.
The overview shown here focuses on the light meson sector, presenting a detailed Partial-Wave Analysis of the processes:
$\pi^- p \to \pi^-\pi^+\pi^- p$ and $\pi^-p\to \pi^-\pi^0\pi^0p$. A new state, the $a_1(1420)$ with $J^{PC}=1^{++}$ is observed.
Its Breit-Wigner parameters are found to be in the ranges: $m = 1412-1422\mev/c^2$ and $\Gamma = 130-150\mev/c^2$. 

In the same analysis, a signal in a wave with $J^{PC}=1^{-+}$ is observed. A resonant origin of this signal 
would not be explicable within the CQM. In addition to this possibility of an exotic state, a
possible non resonant origin of this signal is discussed.


}
\maketitle
\section{The \textsc{Compass} spectrometer}
\label{sec::compass}
The multi-purpose fixed-target spectrometer \compass is located at CERN's northern area and it's supplied with secondary hadron or tertiary muon beams by the Super Proton Synchrotron. 
The two-stage spectrometer setup allows for a wide physics program, which includes e.g. studies of the spin-structure of the nucleon as well as hadron spectroscopy, which will be presented here.\\
For the analysis shown here, data taken in the year 2008 is used. There a $190\gev/c$ negative hadron beam, composed of $\pi^-$ ($97\%$) with some minor contributions from
 $K^-$ ($2\%$) and antiprotons ($1\%$), impinged on a $40\,\mathrm{cm}$ long liquid hydrogen target.\\

The analysis on this data, which is subject to this article, was performed on two three pion final-states, $\pi^-p\to \pi^-\pi^0\pi^0p$ and $\pi^-p\to\pi^-\pi^+\pi^-p$. For these channels, $3.5\cdot 10^6$ and $50\cdot 10^6$ events were recorded in the neutral and charged channel, respectively. The charged data set constitutes up to this point the largest for this particular process.\\
The analyses of both channels were performed independent from each other, using different software packages. The systematic uncertainties of both channels differ, since the reconstructions rely on different parts of the spectrometer.\\ Nevertheless, the physics is the same in both channels, which allows for an effective cross-check of the results.


\begin{figure}[bt]
\centering
\begin{minipage}{0.39\textwidth}
\includegraphics[width=\linewidth]{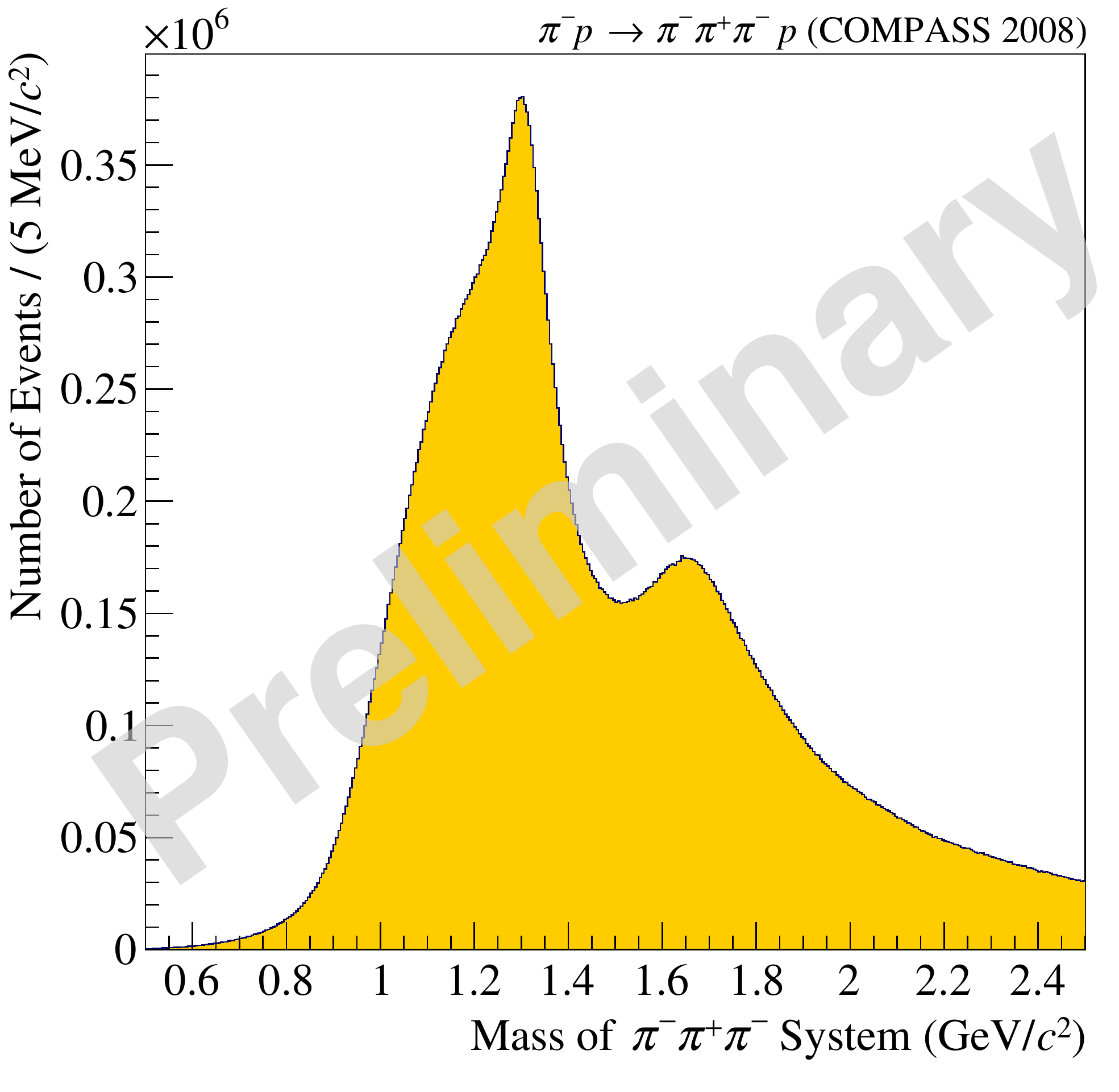}
\caption{Invariant mass distribution of the $\pi^-\pi^+\pi^-$ final state.\vspace*{2\baselineskip}}
\label{pic::specc}
\end{minipage}
\begin{minipage}{0.05\textwidth}
 \hspace{0.05\textwidth}
\end{minipage}
\begin{minipage}{0.49\textwidth}
\centering
\includegraphics[width=\linewidth]{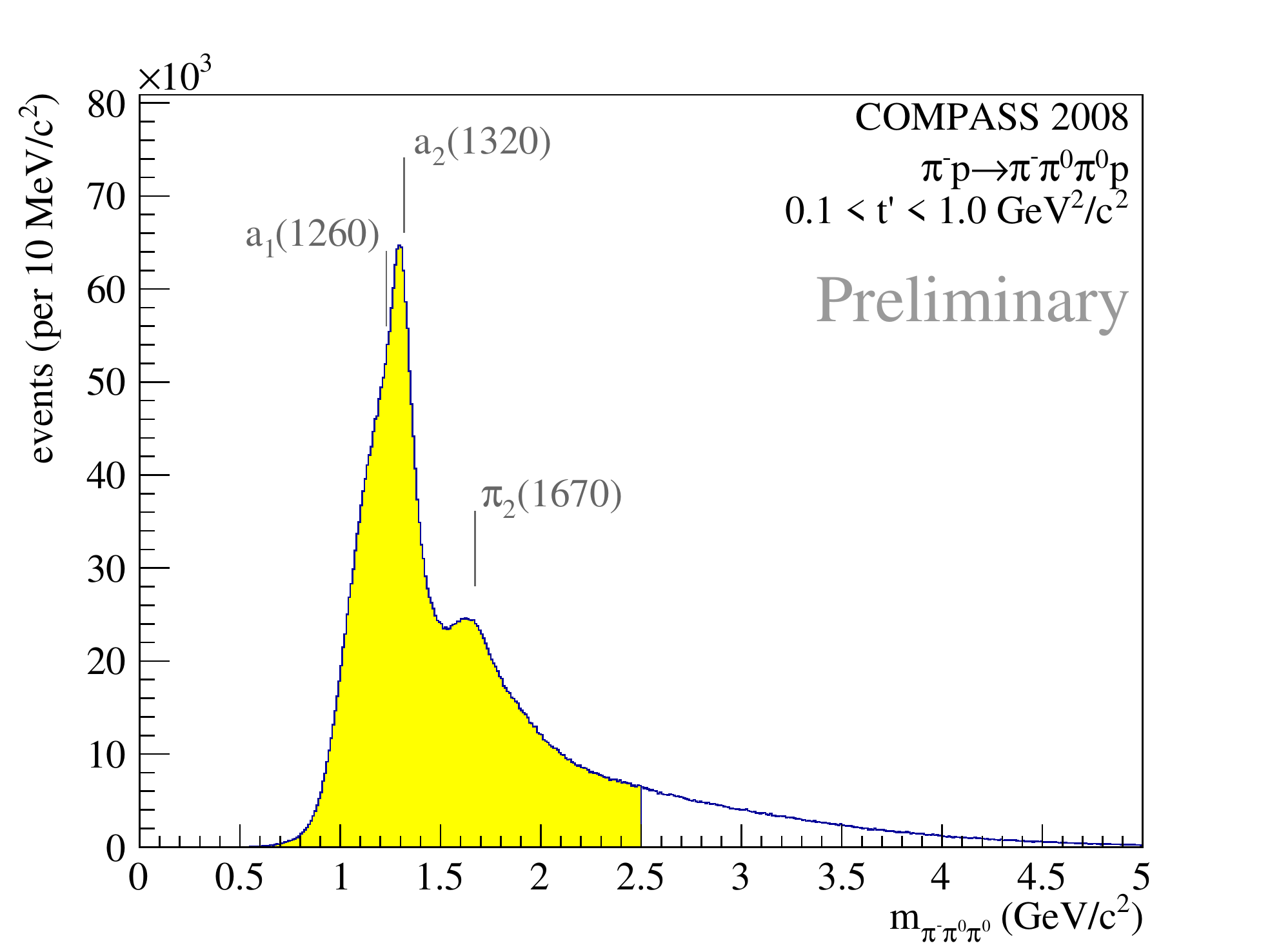}
\caption{Invariant mass distribution of the $\pi^-\pi^0\pi^0$ final state. Labels mark the main resonances described in sec. \ref{sec::major}, wich give the gross features of the spectrum.}
\label{pic::specn}
\end{minipage}
\end{figure}

\section{Analysis method}
\label{sec::pwa}
For the studies presented here, a detailed Partial-Wave Analysis (PWA) was performed on both diffractively produced three-pion final states. In this class of processes, an incoming $\pi^-$ from the beam gets exited via interaction with the target and forms an intermediate state $X^-\!\!$. In our analysis, this interaction is assumed to be dominated by Pomeron exchange. This intermediate state then decays into the observed final-state particles.\\
The excited pion state $X^-$ is characterized by the quantum numbers $J^{PC}M^\epsilon$, where $J$ gives the spin, $P$ and $C$ the eigenvalues of the parity and generalized charge conjugation operators, $M$ the magnetic quantum number and $\epsilon$ the reflectivity.\\
Since the Pomeron itself is not well known, there are many possible values these quantum numbers can take and which all can interfere with each other. Therefore, the main goal of the PWA is to disentangle all appearing contributions of intermediate states with different $J^{PC}M^\epsilon\!$.\\
\begin{figure}[bt]
\centering
\begin{minipage}{0.44\textwidth}
\includegraphics[width=\linewidth]{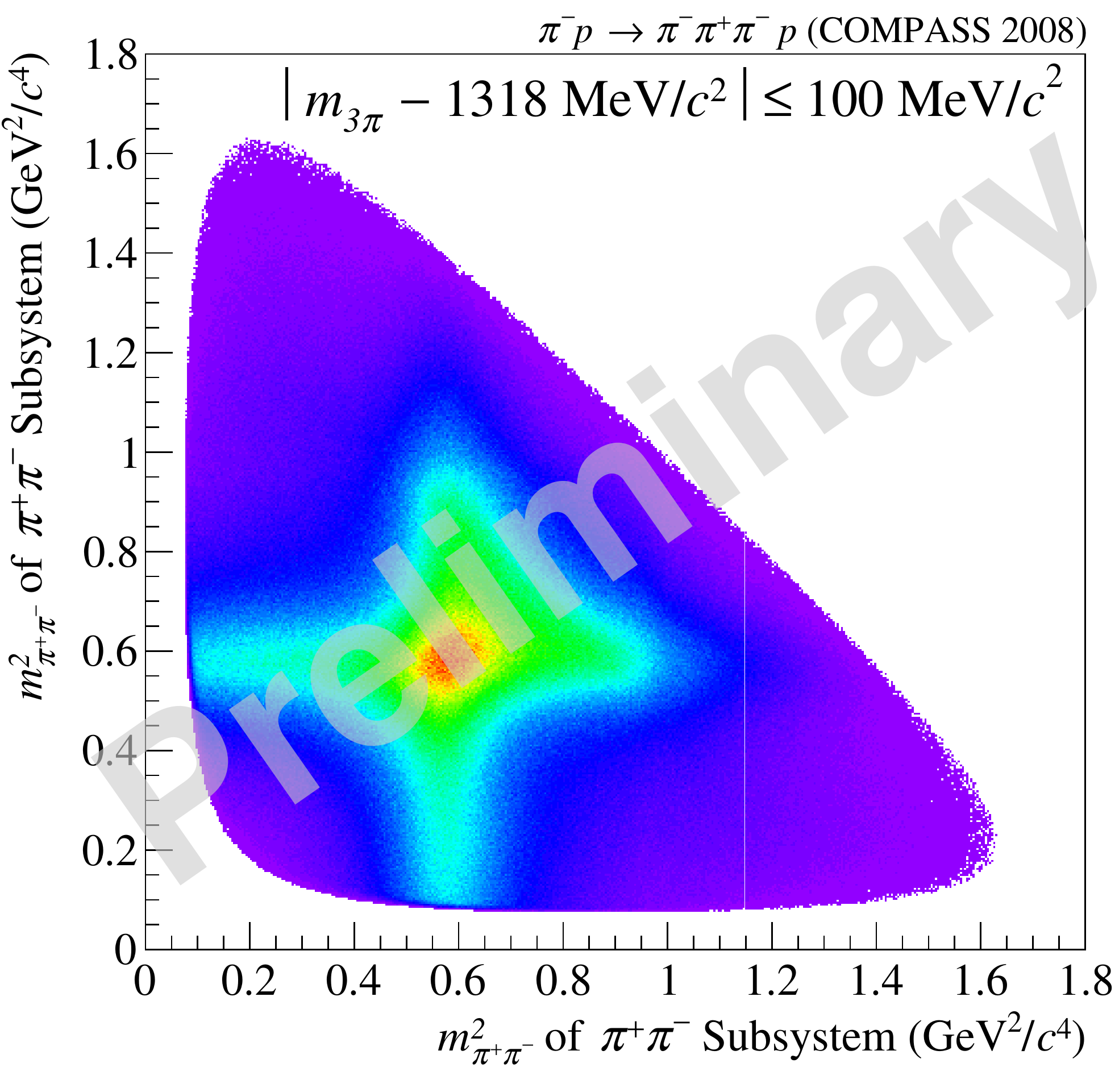}
\caption{Dalitz plot for the $\pi^-\pi^+\pi^-$ channel, with $m_{3\pi}$ chosen around the mass of the $a_2(1320)$}
\label{pic::specc}
\end{minipage}
\begin{minipage}{0.05\textwidth}
 \hspace{0.05\textwidth}
\end{minipage}
\begin{minipage}{0.44\textwidth}
\centering
\includegraphics[width=\linewidth]{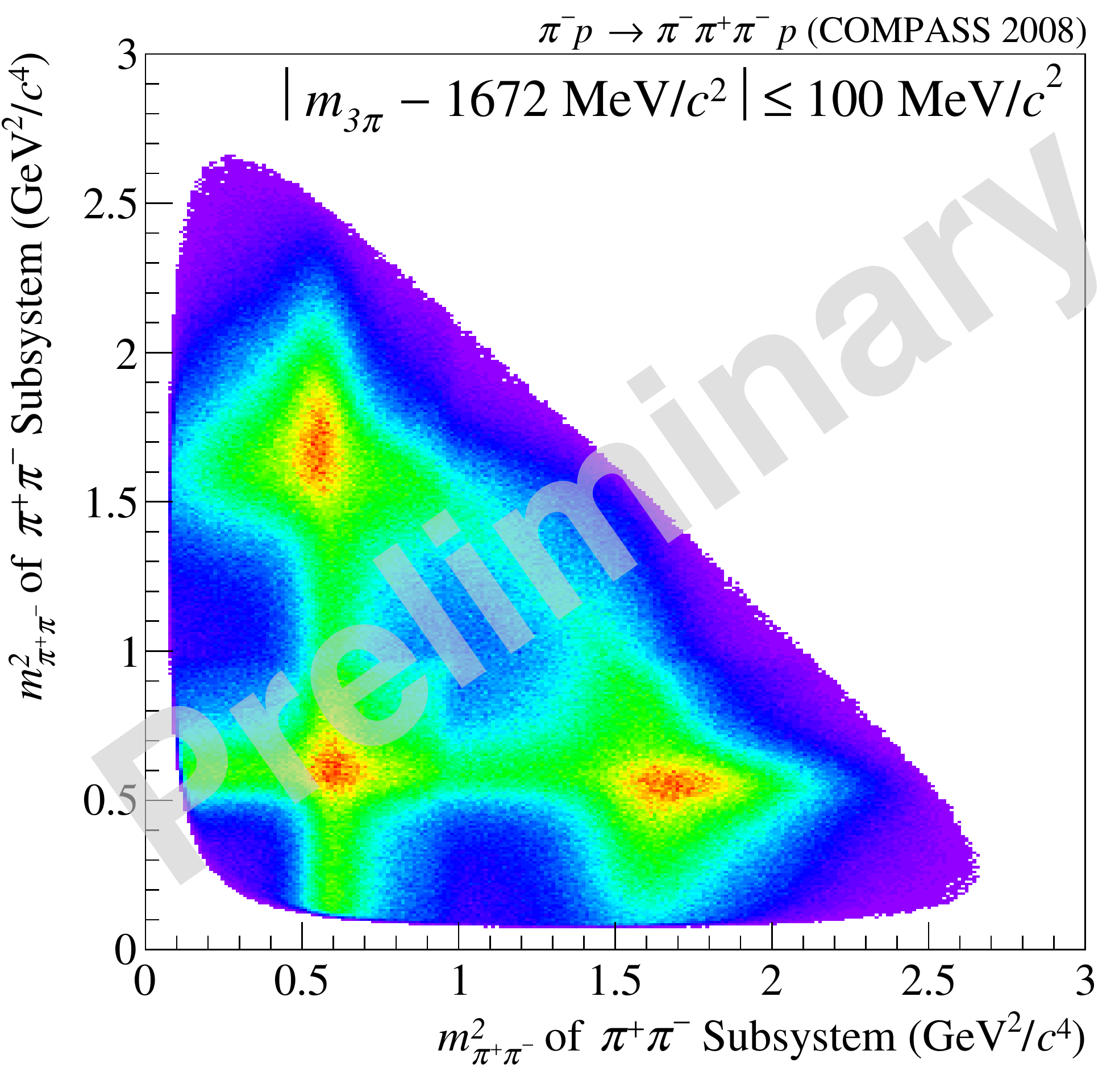}
\caption{Dalitz plot for the $\pi^-\pi^+\pi^-$ channel, with $m_{3\pi}$ chosen around the mass of the $\pi_2(1670)$}
\label{pic::specn}
\end{minipage}
\end{figure}
\subsection{The {isobar}-model}
\label{sec::isobar}
Since the considered processes end up in three-particle final-states, the \textit{isobar} model is employed in the PWA. Here the assumption is made, that the excited states do not decay directly into the final state particles, but perform consecutive two-particle decays into final state particles and/or other intermediate states, the isobars. In the case at hand, $X^-$ decays into a bachelor $\pi$ and an isobar $\xi$, which then decays into the other two pions.\\
An advantage of the the {isobar} model is the factorization of production and decay of $X^-$. Therefore the intensity $\mathcal{I}$, the square of the complex amplitude $\mathcal{A}$, which can be expanded into a series of partial waves, can be written as:
\begin{equation}\label{eq::amp}
\mathcal{I}(m_X,\tau) = \Bigg|\mathcal{A}(m_X,\tau)\Bigg|^2 = \Bigg|\sum_\mathrm{waves} T_\mathrm{wave}(m_X) \psi_\mathrm{wave}(\tau)\Bigg|^2
.\end{equation}
Here, the decay amplitudes $\psi_\mathrm{wave}(\tau)$, depending on the phase-space variables $\tau$, describe the kinematic distribution of the final-state particles, where the production amplitudes $T_\mathrm{wave}$ describe the production of different intermediate states $X^-$ with given $J^{PC}M^\epsilon$\!\!.
Within the isobar model, the decay amplitudes $\psi(\tau)$ can be calculated, putting known complex parametrizations for the line shapes for the isobars into the fit. In our analysis, the following {isobars} were used \cite{Au:1986vs,Beringer:1900zz,Flo:Thesis}:
\begin{center}
 \begin{tabular}{ c|l }
  $I^GJ^{PC}$ &   \\
  \hline
  $ 0^+0^{++}$ & $[\pi\pi]_S,\,f_0(980),\,f_0(1500)$  \\
  $ 1^+1^{--}$ & $\rho(770)$  \\
  $ 0^+2^{++}$ & $f_2(1270)$  \\
  $ 1^+3^{--}$ & $\rho_3(1690)$ \\
 \end{tabular}
\end{center}
With the $\psi(\tau)$ known, the production amplitudes $T(m_X)$ can be extracted from the data by fitting the intensity of eq. (\ref{eq::intens}) to the data in bins of the invariant three-pion mass $m_{3\pi} = m_X$. With this method, no assumptions regarding three-pion resonances have to be made.
\subsection{The wave set}
In the amplitude parametrization of equation (\ref{eq::amp}) a sum over a given set of $\mathrm{waves}$ appears, which are defined by:
\begin{equation}
J^{PC}M^\epsilon [\mathrm{isobar}]\ \pi\ L
,\end{equation}
where $J^{PC}M^\epsilon$ give the quantum numbers of the state $X^-$\!\!, while the rest describes its decay mode. Since the {isobars} are all well-known states, their quantum numbers are not explicitly stated in the formula above. $L$ is the relative orbital angular momentum between the {isobar} and the bachelor pion.\\
In the analyses presented, a set of $87$ waves with a spin $J$ and angular momentum $L$ up to six was used. In addition, one incoherent isotropic wave was added in order to be able to describe uncorrelated events. \cite{Flo:Thesis}
\label{sec::waveset}

\section{Selected results}
\label{sec::results}
\subsection{The biggest waves}
\label{sec::major}
\begin{figure}[bt]
\centering
\begin{minipage}{0.44\textwidth}
\includegraphics[width=\linewidth]{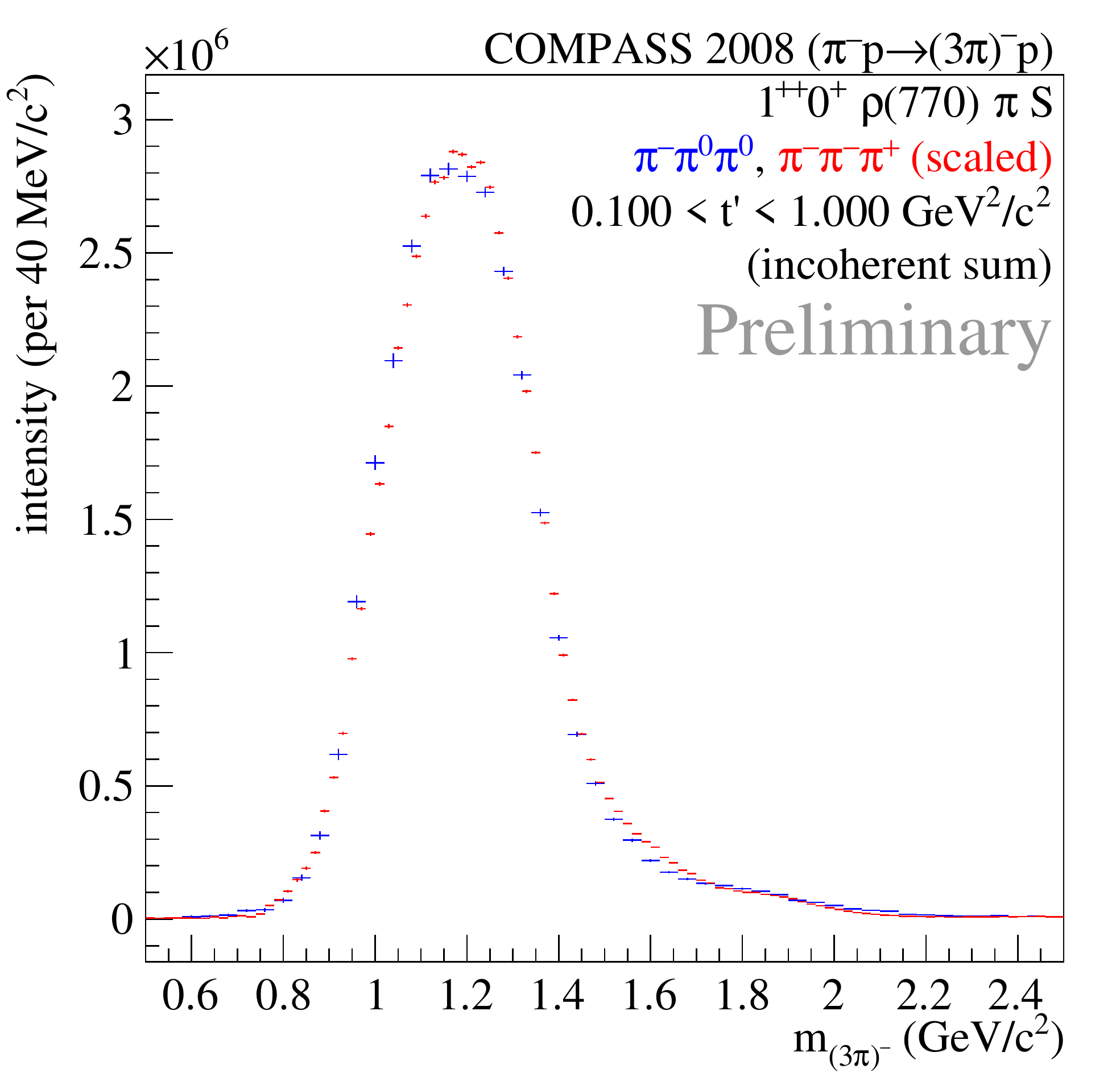}
\caption{Intensity of the $1^{++}0^+\rho(770)\ \pi\ S$ wave for both three-pion channels. The intensity of the charged channel is scaled to match the integral of the neutral channel.}
\label{pic::1pp}
\end{minipage}
\begin{minipage}{0.05\textwidth}
 \hspace{0.05\textwidth}
\end{minipage}
\begin{minipage}{0.44\textwidth}
\centering
\includegraphics[width=\linewidth]{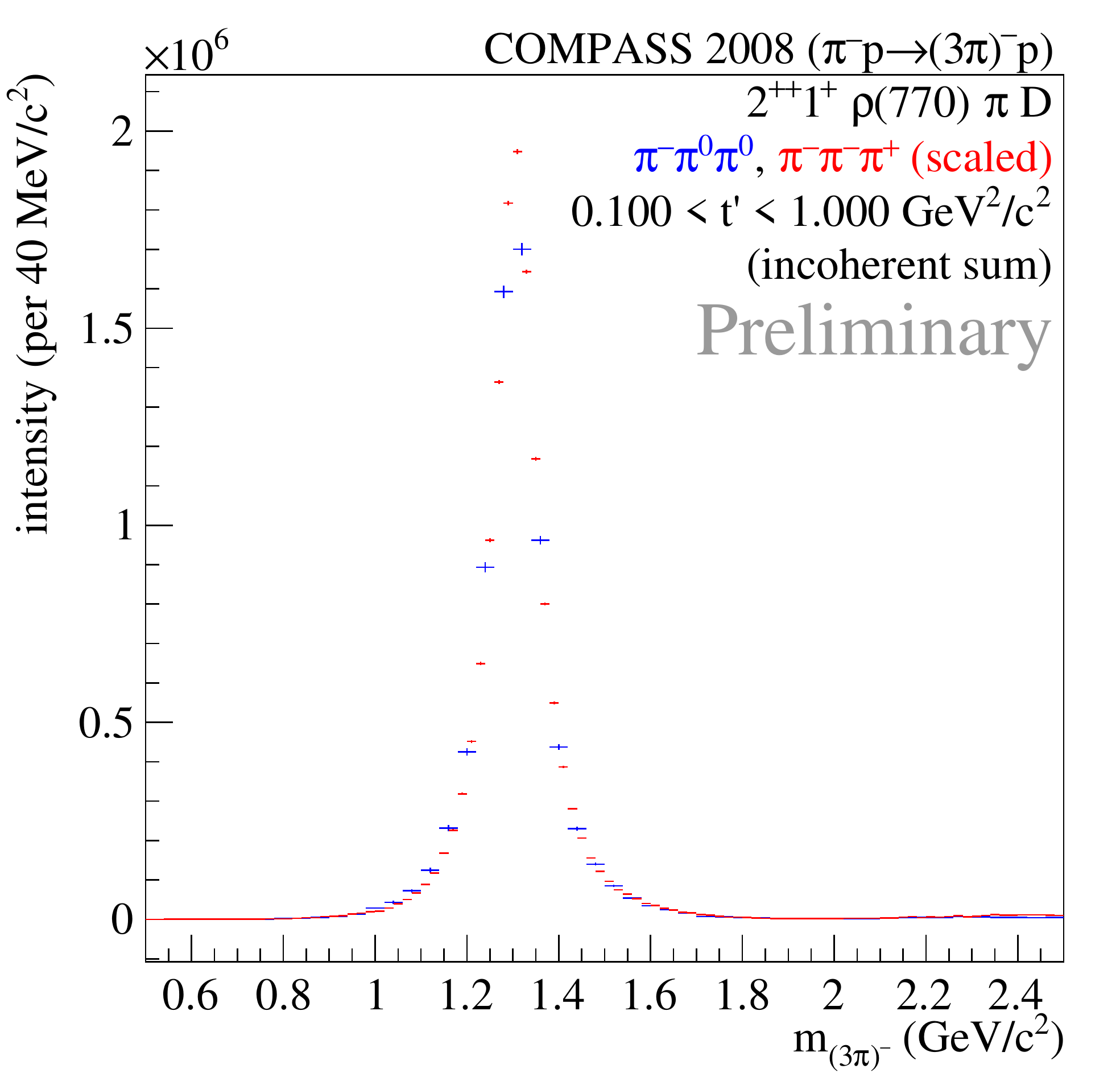}
\caption{Intensity of the $2^{++}1^+\rho(770)\ \pi\ D$ wave for both three-pion channels. The intensity of the charged channel is scaled to match the integral of the neutral channel.}
\label{pic::2pp}
\end{minipage}
\end{figure}
\begin{figure}[bt]
\centering
\begin{minipage}{0.44\textwidth}
\includegraphics[width=\linewidth]{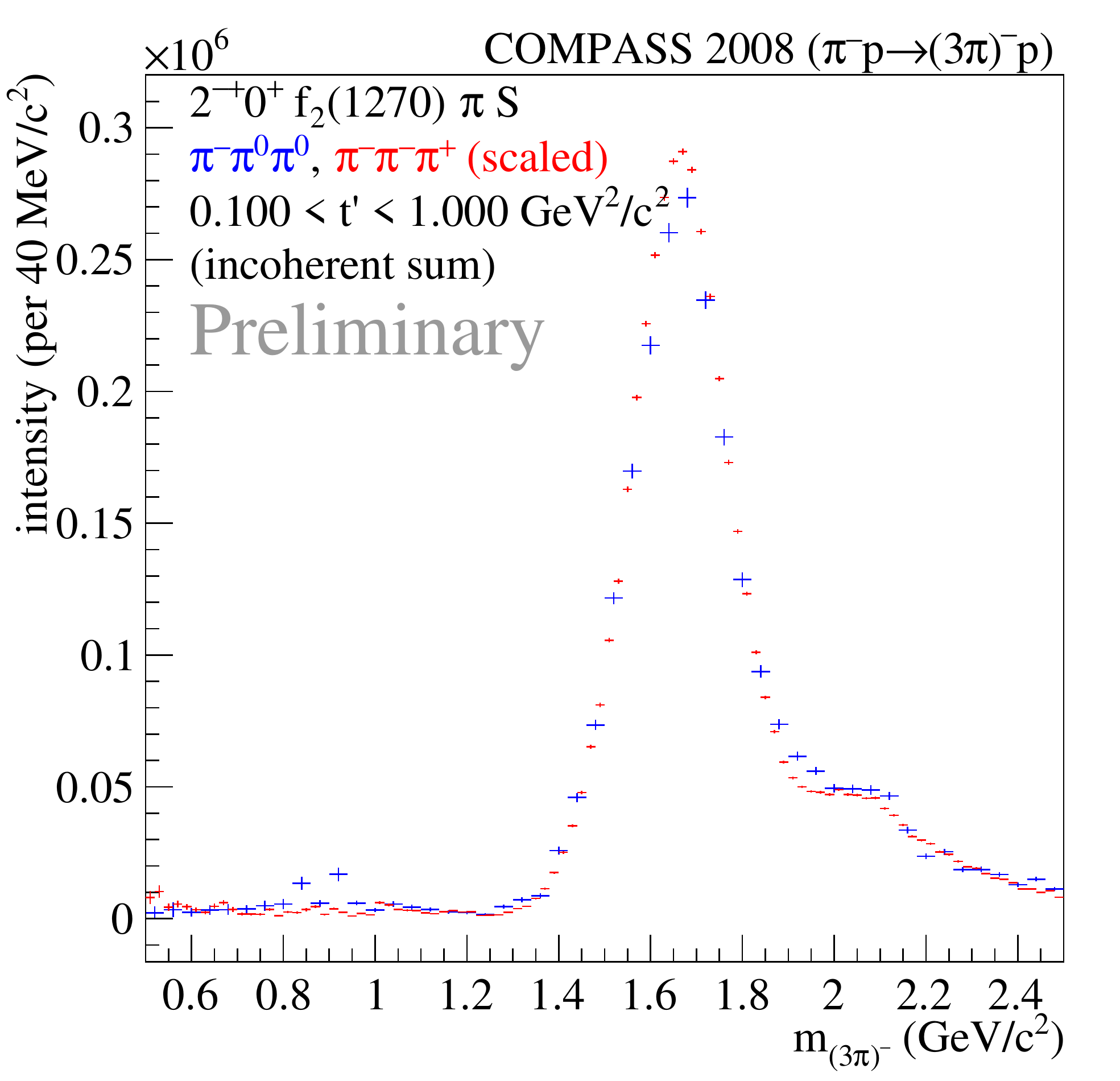}
\caption{Intensity of the $2^{-+}0^+f_2(1270)\ \pi\ S$ wave for both three-pion channels. The intensity of the charged channel is scaled to match the integral of the neutral channel.}
\label{pic::2mp}
\end{minipage}
\begin{minipage}{0.05\textwidth}
 \hspace{0.05\textwidth}
\end{minipage}
\begin{minipage}{0.44\textwidth}
\centering
\includegraphics[width=\linewidth]{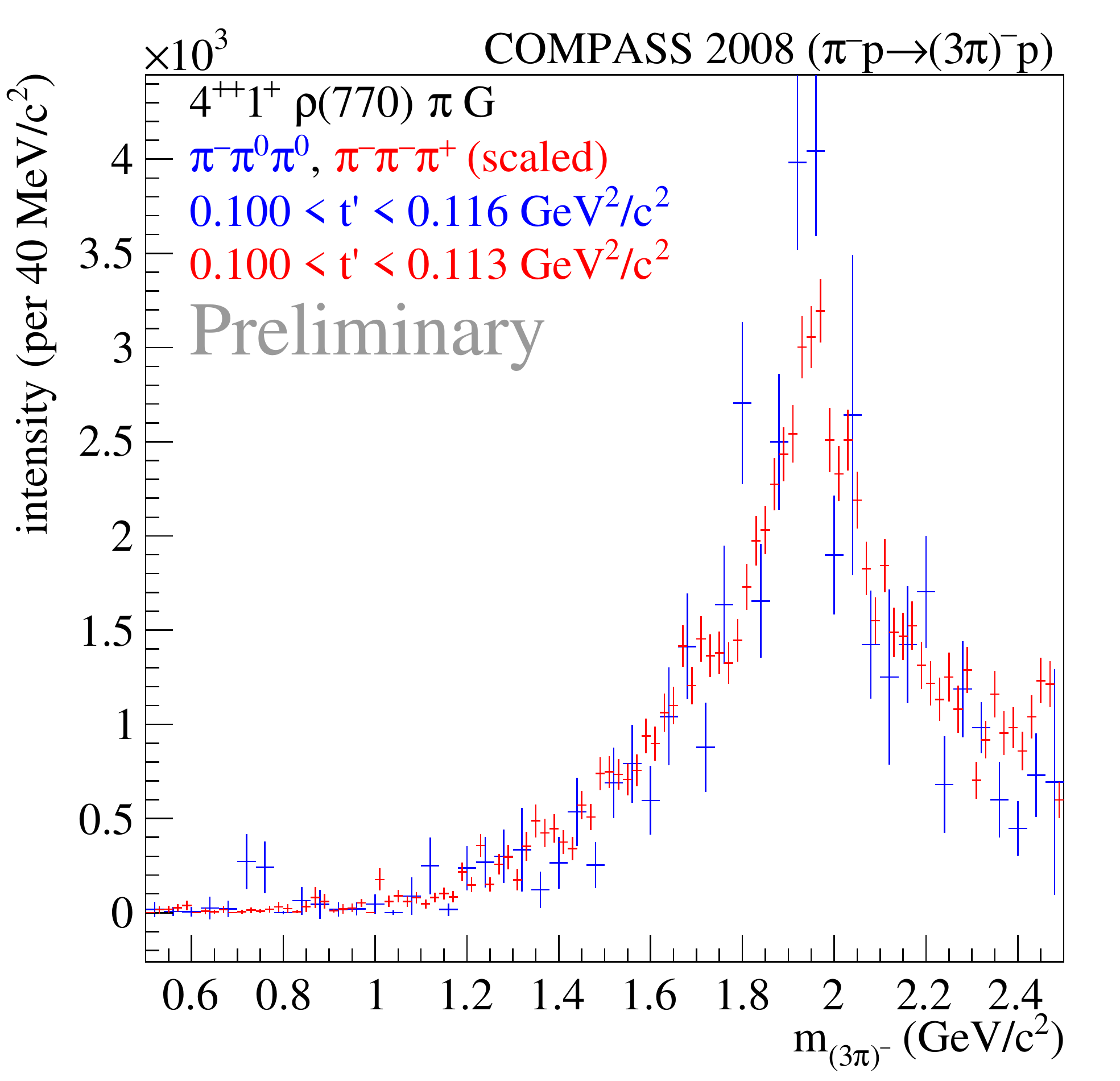}
\caption{Intensity of the $4^{++}1^+\rho(770)\ \pi\ G$ wave for both three-pion channels.  The intensity of the charged channel is scaled to match the integral of the neutral channel.}
\label{pic::4pp}
\end{minipage}
\end{figure}
The main features of the invariant mass spectra of both channels, shown in Figs. \ref{pic::specc} and \ref{pic::specn}, can be explained by the three biggest waves in the analysis, these are:
\begin{itemize}
\item $1^{++}0^+\rho(770)\ \pi\ S$, Fig. \ref{pic::1pp}: This wave describes an axial-vector intermediate state, decaying into $\rho(770)$ and a pion. With $33\%$ of the intensity in the charged channel, this wave is the biggest in the analysis. The dominant structure visible is the $a_1(1260)$ resonance. A good agreement between neutral an charged channel can be seen.
\item $2^{++}1^+\rho(770)\ \pi\ D$, Fig. \ref{pic::2pp}: This wave is the second biggest wave in the analysis, taking $8\%$ of the total intensity. It describes a spin-2 meson, which also decays into $\rho(770)\ \pi$. In this wave, the clearest three-pion resonance, the $a_2(1320)$ can be seen, again with a good agreement between both channels.
\item $2^{-+}0^+f_2(1270)\ \pi\ S$, Fig. \ref{pic::2mp}: The third biggest wave gives rise to the second peak in the invariant mass-spectra. It describes an intermediate state behaving like a pion with spin 2 decaying into $f_2(1270)\ \pi$. The main resonance visible is the $\pi_2(1670)$. This wave takes $7\%$ of the charged channel's intensity.
\end{itemize}
The Partial-Wave Analysis performed is not just able to extract the gross features of the spectra, but can resolve contributions down to the sub-percent level. One of these waves is $4^{++}1^+\rho(770)\ \pi\ G$, depicted in Fig. \ref{pic::4pp}, which describes a spin 4 meson decaying $\rho(770)\ \pi$. This wave takes only about $0.76\%$ intensity, but the $a_4(2040)$ resonance can be clearly seen in both channels.
\subsection{The $\mathbf{a_1(1420)}$}
\label{sec::a1}
\begin{figure}[bt]
\centering
\begin{minipage}{0.44\textwidth}
\includegraphics[width=\linewidth]{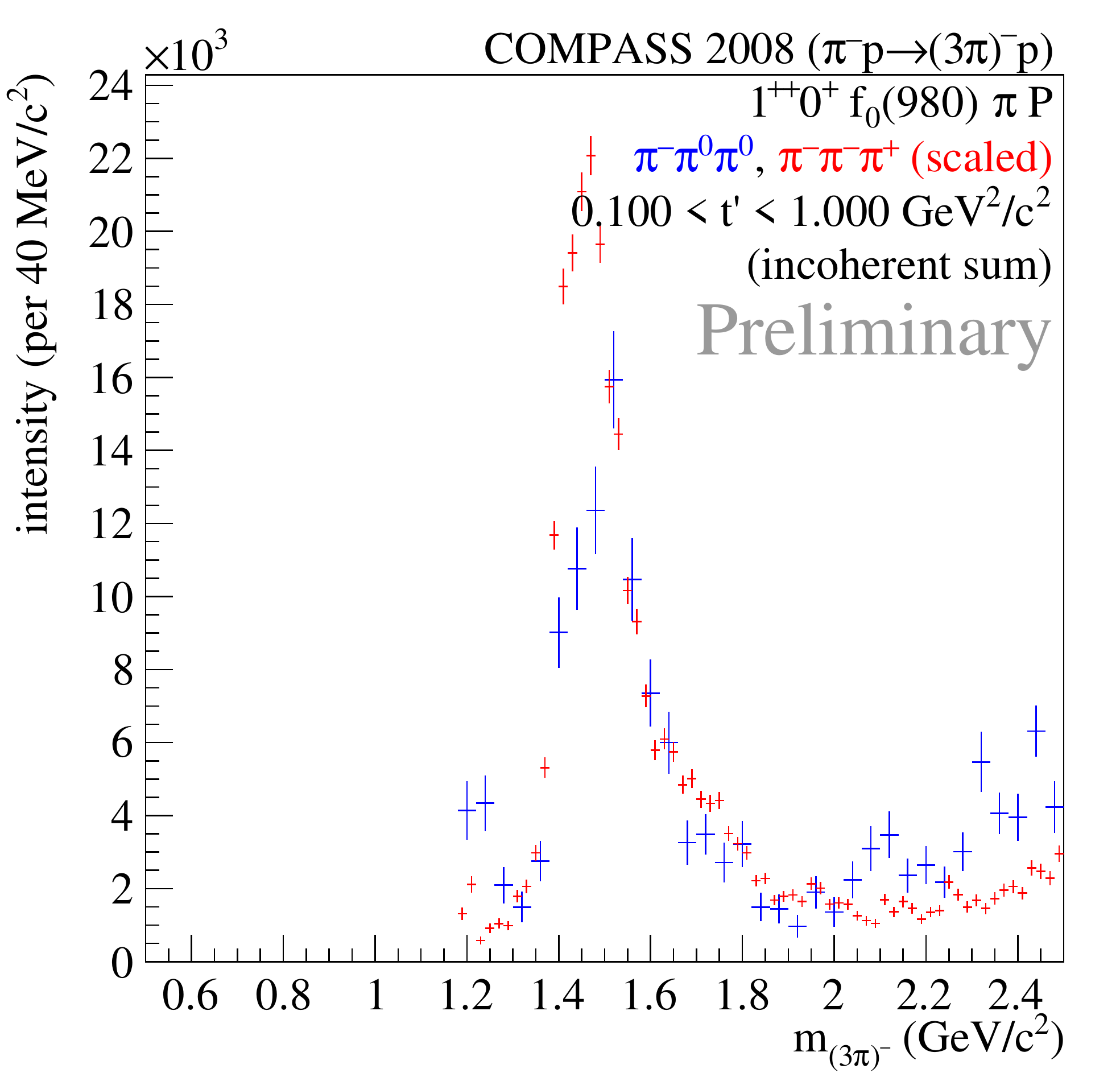}
\caption{Intensity of the $1^{++}0^+f_0(980)\ \pi\ P$ wave. The peak of the $a_1(1420)$ is clearly visible in both channels.\vspace*{\baselineskip}}
\label{pic::1420}
\end{minipage}
\begin{minipage}{0.05\textwidth}
 \hspace{0.05\textwidth}
\end{minipage}
\begin{minipage}{0.44\textwidth}
\centering
\includegraphics[width=\linewidth]{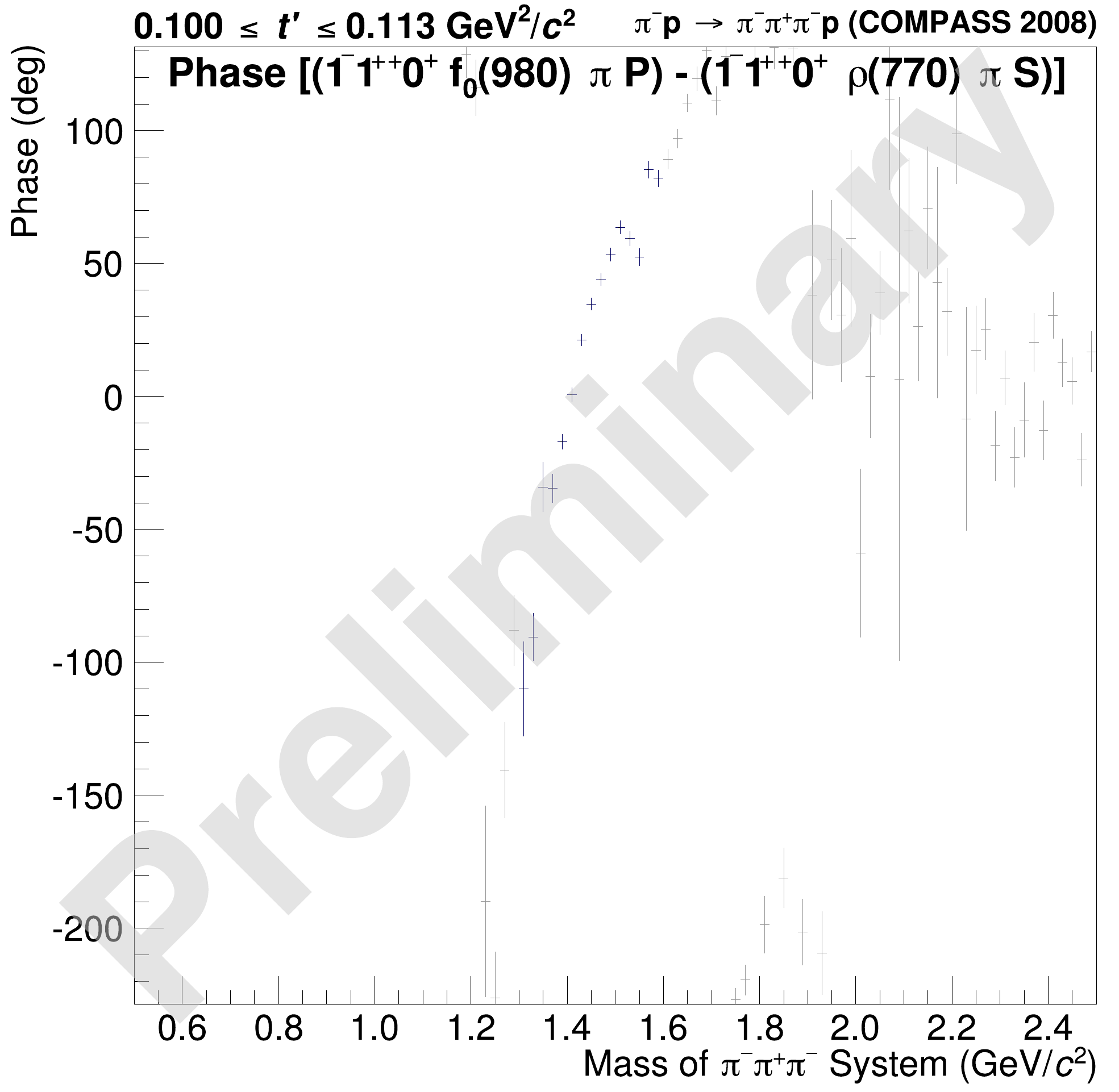}
\caption{Relative phase between the $1^{++}0^+\rho(770)\ \pi\ S $ and the $1^{++}0^+f_0(980)\ \pi\ P$ waves. In the mass region of the $a_1(1420)$, a clear phase motion is visible.}
\label{pic::1260:1420}
\end{minipage}
\end{figure}
Besides the well know resonances, shown up to now, a new resonance, the $a_1(1420)$ was seen in the $1^{++}0^+f_0(980)\ \pi\ P$ wave. The corresponding intermediate state has the same quantum numbers as the biggest wave, but it decays via a peculiar decay mode, $f_0(980)\ \pi$.\\
This new resonance appears in both channels, shown in Fig. \ref{pic::1420}. A clear and rapid phase motion can be seen with respect to the biggest wave in the model, depicted in Fig. \ref{pic::1260:1420}. This indicates, that the observed intensity peak constitutes an actual resonance. The extracted Breit-Wigner parameters, mass and width, lie in the following ranges:
\begin{eqnarray}
m = 1412-1422\mev/c^2\ \ \ \ \ \ \ \ \ \ \ \ \ \ \ \Gamma = 130-150\mev/c^2
\end{eqnarray}
Observing a new resonance in this mass region interesting for a number of reasons. First, no model or lattice QCD predicted a resonance in said mass region. Second, the $a_1(1420)$ decays into $f_0(980)\ \pi$ with an unusually small relative intensity compared to other resonances. Additionally, the $f_0(980)$ is also known to strongly couple $KK$ states. Third, the mass of the $a_1(1420)$ is only slightly bigger than the $KK^*$ threshold. This might be a hint for a dynamic nature of this resonance involving $KK^*$ loop diagrams, but the true nature of the $a_1(1420)$ remains unclear at the moment. \cite{Basdevant:1977ya}
\subsection{Spin exotic signal}
\label{sec::exotic}
In addition to the waves shown previously, which all had non-exotic quantum numbers, i.e. they can be explained within
the Constituent-Quark Model (CQM), a signal has been observed in the $1^{-+}1^+\rho(770)\,\pi\,P$ wave.
Due to its exotic quantum numbers $J^{PC} = 1^{-+}$, thus behaving like a pion with spin 1, a resonance in this wave would exceed the scope of the CQM. As in the previous waves, the signal is seen in the charged, as well as the 
neutral channel.
\begin{figure}[bt]
\centering
\begin{minipage}{0.44\textwidth}
\includegraphics[width=\linewidth]{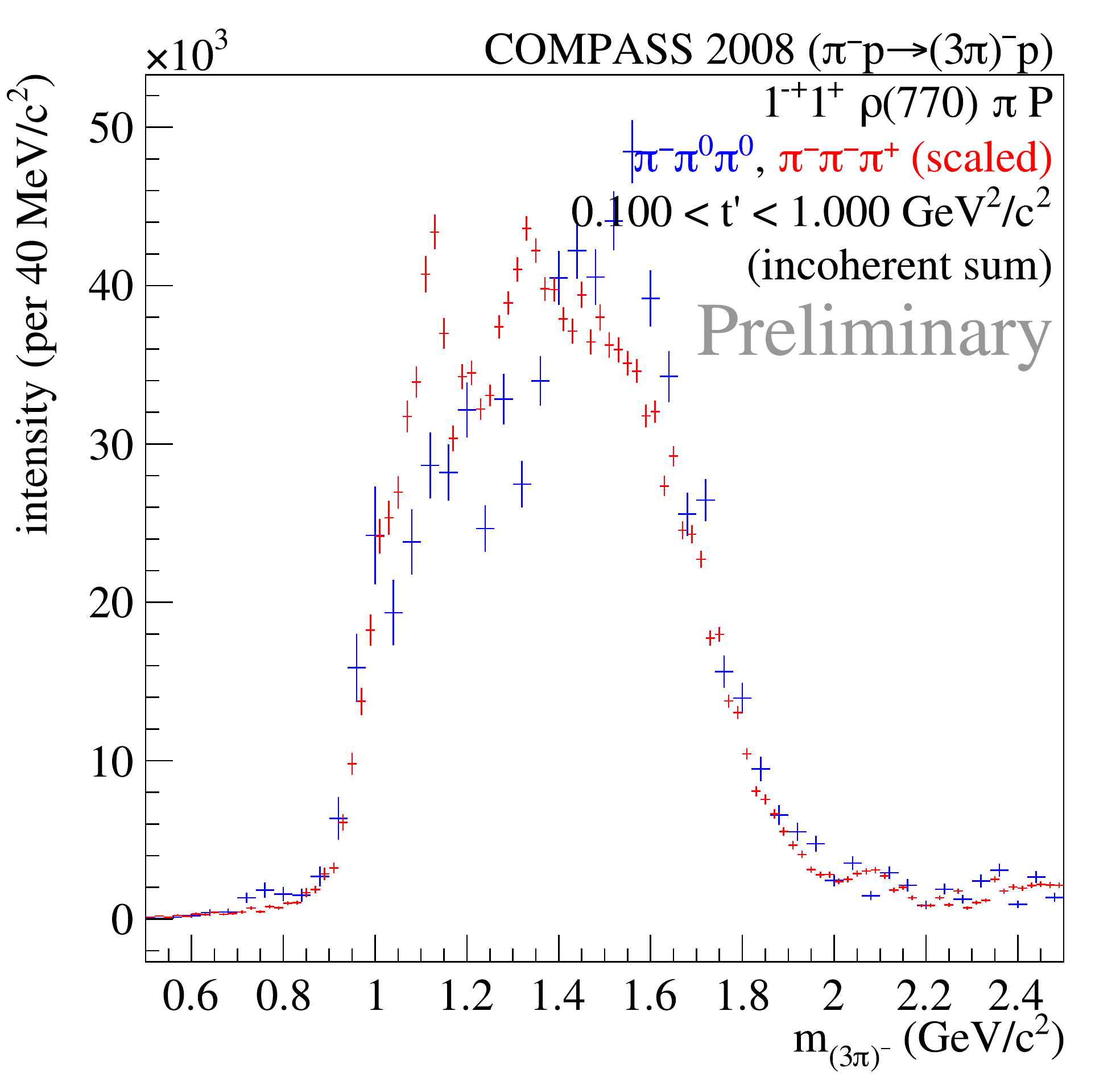}
\caption{Intensity of the psin exotic $1^{-+}1^+\rho(770)\ \pi\ P$ wave. A similar signal is visible in both cahnnels.\vspace*{2\baselineskip}}
\label{fig::exitic_intens}
\end{minipage}
\begin{minipage}{0.05\textwidth}
 \hspace{0.05\textwidth}
\end{minipage}
\begin{minipage}{0.44\textwidth}
\centering
\includegraphics[width=\linewidth]{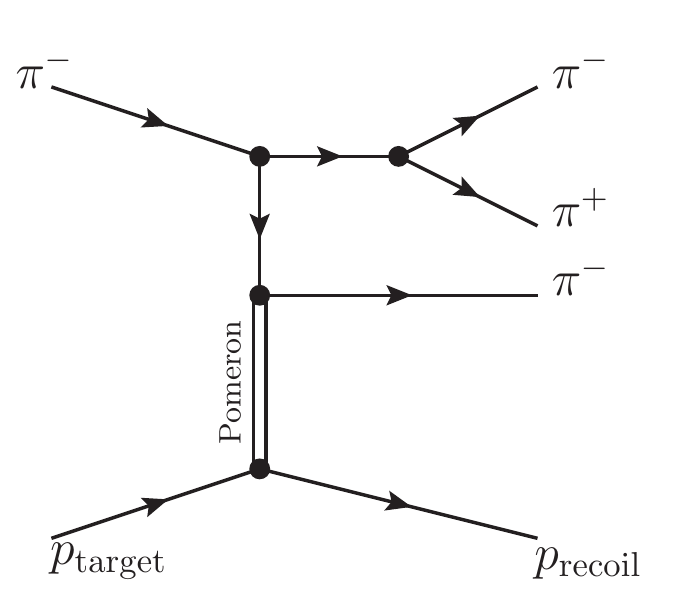}
\caption{Diagrammatic view of the non-resonant Deck effect. In this process, no three-pion intemediate state appears.}
\label{fig::deck}
\end{minipage}
\end{figure}
\\Besides a possible resonance, there exist other, non-resonant, mechanisms that can lead to a signal in said wave.
One of these mechanisms is the so-called Deck effect, which is diagrammatically shown in fig. \ref{fig::deck}. To study its effect, 
a PWA was performed on pseudo-data generated with a Deck model for the charged channel. The results of the Deck's Partial-Wave decomposition are then compared with the results 
for real data.\\
To further disentangle such effects, this study is also performed in bins of $t'$, with:
\begin{equation}
\label{eq::tprime}
t' = |t| - |t_\mathrm{min}| \simeq -t
\end{equation}
where $t$ is the four-momentum transfer between beam pion and target proton. This allows to better disentangle resonant and non-resonant contributions since the resonance parameters may not depend on $t'$.\\
\begin{figure}[bt]
\centering
\begin{minipage}{0.44\textwidth}
\includegraphics[width=\linewidth]{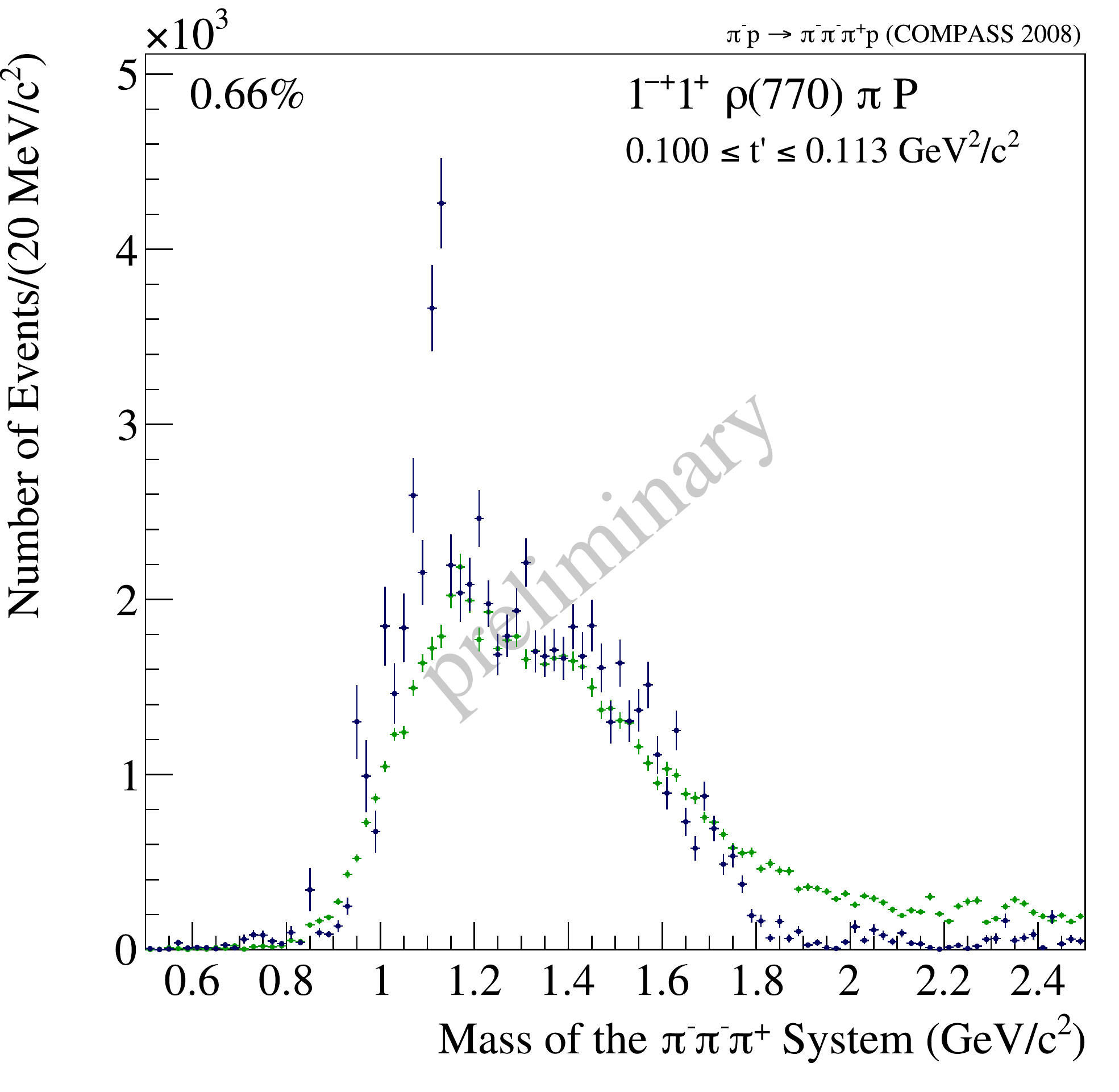}
\end{minipage}
\begin{minipage}{0.05\textwidth}
 \hspace{0.05\textwidth}
\end{minipage}
\begin{minipage}{0.44\textwidth}
\centering
\includegraphics[width=\linewidth]{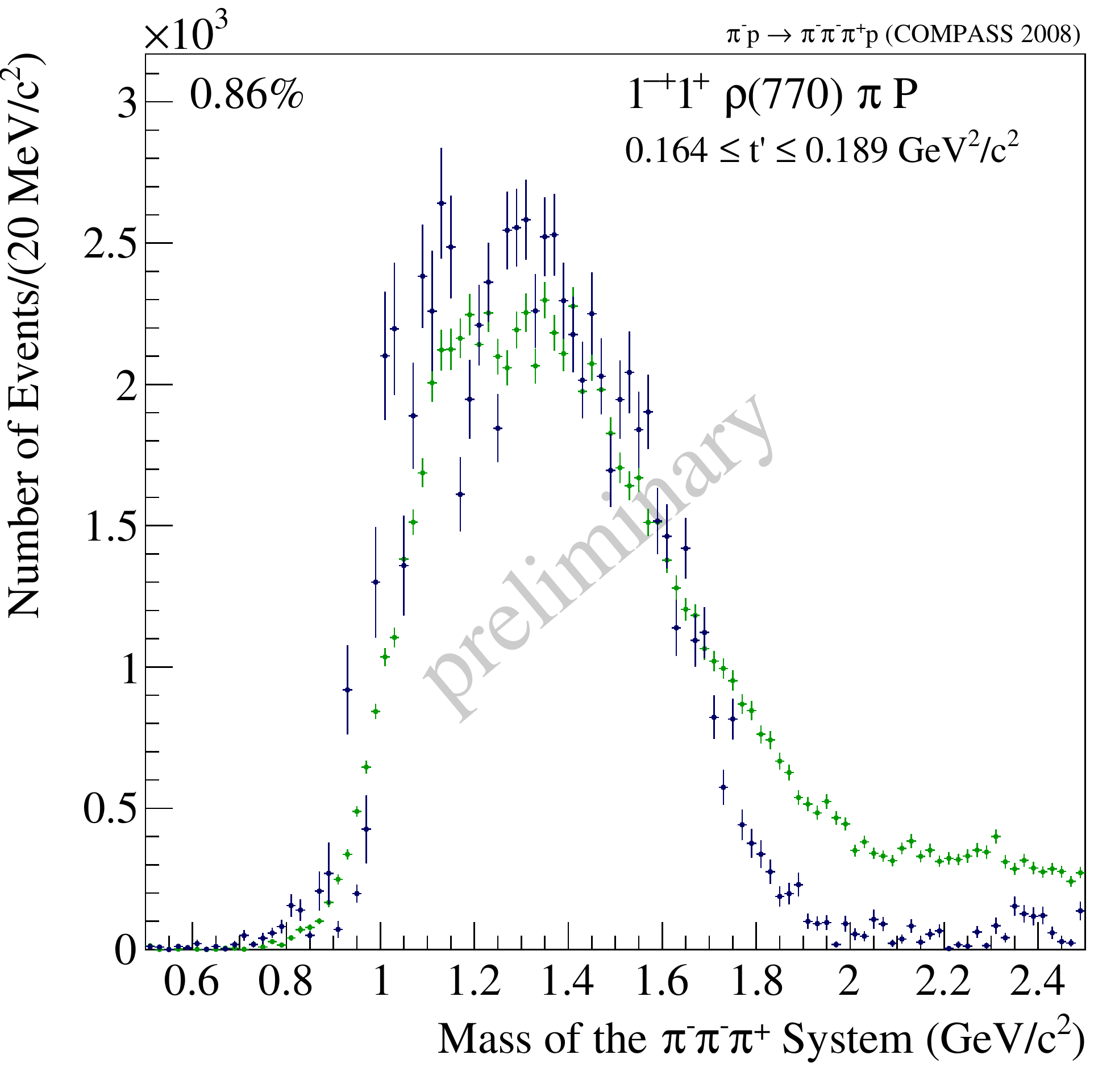}
\end{minipage}
\caption{Intensity distribution for the spin exotic wave for the data (blue) and the Deck Monte-Carlo (green), normalized to the integrals over mass and $t'$, at low values of the four momentum transfer. The intensity in the data can be explained well by the Deck effect.}
\label{fig::exotic_low_t}
\end{figure}
\begin{figure}[bt]
\centering
\begin{minipage}{0.44\textwidth}
\includegraphics[width=\linewidth]{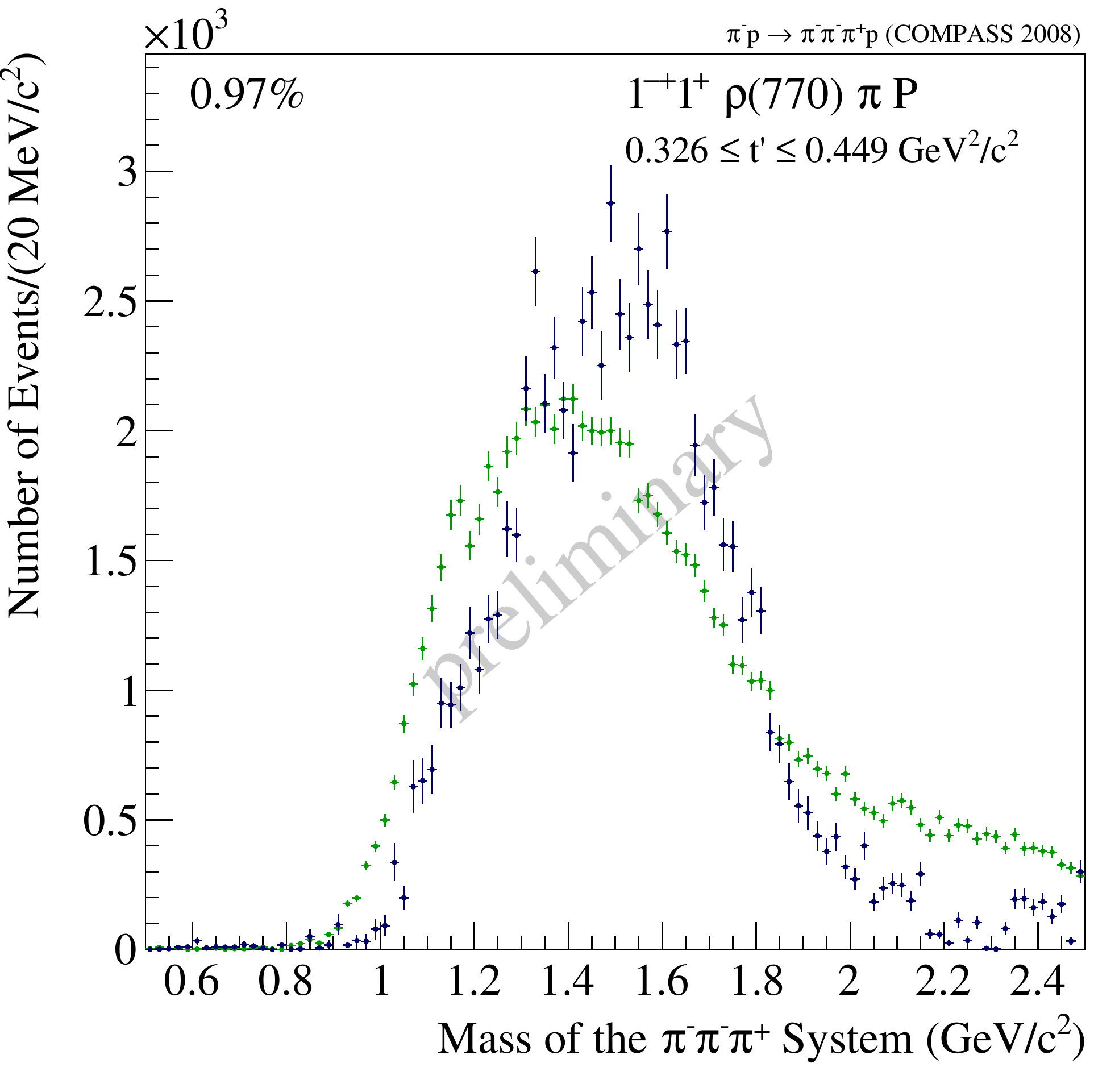}
\end{minipage}
\begin{minipage}{0.05\textwidth}
 \hspace{0.05\textwidth}
\end{minipage}
\begin{minipage}{0.44\textwidth}
\centering
\includegraphics[width=\linewidth]{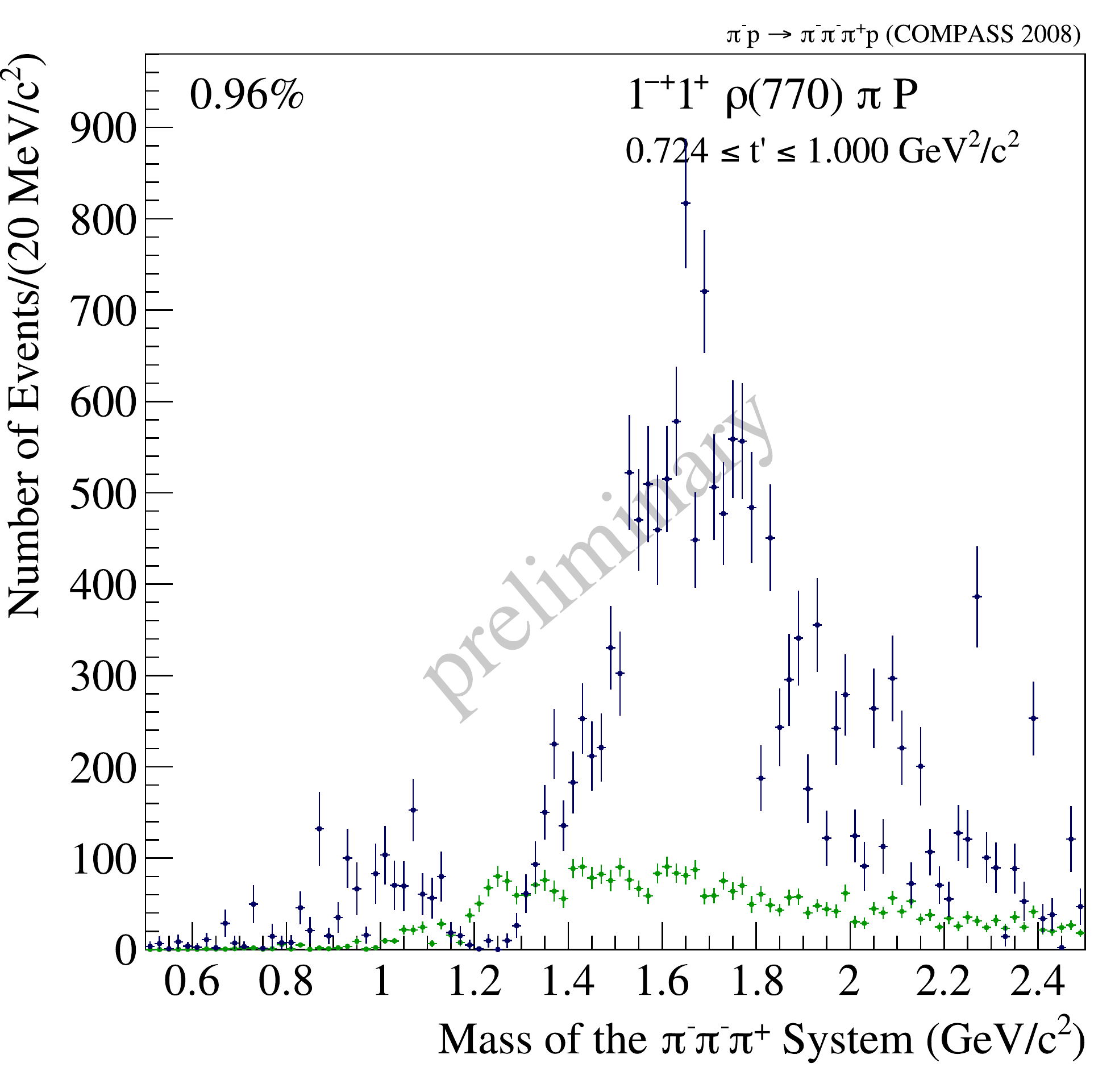}
\end{minipage}
\caption{Intensity distribution for the spin exotic wave for the data (blue) and the Deck Monte-Carlo (green), normalized to the integrals over mass and $t'$, at high values of the four momentum transfer. The Deck contributions vanishes, while the signal stays also at high $t'$.}
\label{fig::exotic_high_t}
\end{figure}

\section{Conclusions}
\label{sec::conclusion}
Due to the large three pion data sets collected by \compass, a very detailed Partial-Wave Analysis could be performed, allowing for a systematic cross check between the two channels and thus giving a deep insight into the spectrum of light mesons.\\
For the present analysis a large wave-set, consisting of $87$ waves up to spin 6 was employed, and waves contributing only on the sub-percent level to the total intensity could be extracted.
Besides reproducing all well known resonances with isospin $I^G = 1^-$, a previously unknown state, the $a_1(1420)$, could be extracted in the $1^{++}0^+f_0(980)\ \pi\ P$ wave.\\
In addition to these resonances, which all have quantum numbers that can be explained by the CQM, a signal in the spin exotic $1^{-+}1^+\rho(770)\,\pi\,P$ wave was seen. A possible, non-resonant origin of this signal, the Deck effect, was studied and compared to the data. With the Deck-effect, the signal corresponding to low four-momentum transfer can be explained well, while for high $t'$, it shows an excess over the Deck-decomposition.

\section{Outlook}
\label{sec::outlook}
Since the analyses presented here did not involve any line shape for the resonances, mass dependent fits have to be performed, to determine these and thus extracting masses and widths of the appearing resonances.\cite{Proceeding:Haas}\\
The presented analysis still relies on the isobar model, which assumes fixed line-shapes for the appearing isobars. To see, if this assumption is justified, a new method is being studied to directly extract the isobar shapes from the data. This allows to check the validity of the isobar model and to determine resonance parameters for the isobars as well.\\
Due to the large number of events, also non-resonant contributions become important. One of them, the Deck contribution introduced above, is expected to have a big impact on the data. To study its particular influence on the PWA will be the aim of further studies.\cite{Fabi:Bormio}


\end{document}